\documentclass[a4paper]{PoS}
\usepackage{amsmath}
\usepackage{amssymb}
\usepackage{pifont}
\usepackage{graphicx}
\usepackage{url}

\title{Automating Predictions for Standard Model Effective Field Theory in
	{\sc MadGraph5\_aMC@NLO}}

\ShortTitle{EFT with {\sc MG5\_aMC@NLO}}

\author{\speaker{Cen Zhang}\\
        Department of Physics, Brookhaven National Laboratory\\
	Upton, NY 11973, USA\\
        E-mail: \email{cenzhang@bnl.gov}}

\abstract{
Next-to-leading order event generation for the Standard Model effective field
theory has started to become available in the {\sc MadGraph5\_aMC@NLO}
framework.  In this talk we discuss some of the recent progresses in this
direction, with a focus on the top-quark sector.
}

\FullConference{12th International Symposium on Radiative Corrections (Radcor
2015) and LoopFest XIV (Radiative Corrections for the LHC and Future
Colliders)\\ 15-19 June, 2015\\ UCLA Department of Physics \& Astronomy Los
Angeles, USA}

\begin{document}

\section{Introduction}
The Standard Model (SM) effective field theory (EFT) at
dimension-six \cite{Buchmuller:1985jz} is a powerful approach to
searching for SM deviations through precision measurements of the SM. By
supplementing the SM Lagrangian with a set of higher-dimensional operators,
indirect effects of heavy particles, possibly beyond the reach of the LHC,
can be consistently accommodated.  The
program of determining the SM Lagrangian up to dimension-six has been
successful at the LHC Run-I \cite{Agashe:2014kda}, and will continue to proceed
at Run-II with improved precisions.  To this end, accurate predictions from the
theory side are needed, not only for the SM contribution but also for the
deviations, in order to set the most stringent bounds on new physics scales and
also to better characterize the signals and improve the sensitivities.  This
motivates us to study the higher-order corrections within the SM EFT, and in
particular, the next-to-leading order (NLO) QCD corrections are usually
the most important ones at the LHC.

Recently, NLO simulation for the SM EFT is becoming available in the {\sc
MadGraph5\_aMC@NLO} ({\sc MG5\_aMC}) framework \cite{Alwall:2014hca}.  The
framework features a fully automatic approach to NLO QCD calculation interfaced
with shower via the {\sc MC@NLO} method \cite{Frixione:2002ik}.  The dimension-six
Lagrangian can be implemented with the {\sc FeynRules} package \cite{Alloul:2013bka},
whose output is in the UFO format \cite{Degrande:2011ua} and can be directly
passed to {\sc MG5\_aMC} for event generation.  Loop corrections are computed
by {\sc MadLoop} \cite{Hirschi:2011pa}, where the ultraviolet and the $R_2$
counterterms are required and can be computed by the {\sc NLOCT} package
\cite{Degrande:2014vpa}, up to dimension-four.  The {\sc NLOCT} package is
currently being developed to also cover the dimension-six counterterms from
effective vertices.  In the meantime, case studies can be done by considering
a certain set of operators and processes, and computing the relevant
counterterms separately.  Once the NLO UFO model is made, dimension-six
contributions to the cross sections and differential distributions can be
computed up to NLO in QCD, in a fully automatic way like in the SM.  In this
talk we summarize some of these studies, with a focus on the top-quark sector.

\section{The top-quark sector}
\newcommand{\FDF}{\frac{y_t^2}{2}\left(\varphi^\dagger\overleftrightarrow{D}_\mu\varphi\right)}
\newcommand{\FDFI}{\frac{y_t^2}{2}\left(\varphi^\dagger\overleftrightarrow{D}^I_\mu\varphi\right)}

We start with the flavor diagonal interactions in the top-quark sector.  To
probe deviations in the top-quark couplings, a number of dimension-six
operators involving two quark fields in the third generation are relevant
\cite{AguilarSaavedra:2008zc,AguilarSaavedra:2009mx}:
\begin{equation}
  \begin{array}{ll}
    O_{\varphi Q}^{(3)}=i\FDFI(\bar{Q}\gamma^\mu\tau^IQ),
    &
    O_{\varphi Q}^{(1)}=i\FDF(\bar{Q}\gamma^\mu Q),
    \\
    O_{\varphi t}=i\FDF(\bar{t}\gamma^\mu t),
    &
    O_{tB}=g_Yy_t(\bar{Q}\sigma^{\mu\nu}t)\tilde{\varphi}B_{\mu\nu},
    \\
    O_{tW}=g_Wy_t(\bar{Q}\sigma^{\mu\nu}\tau^It)\tilde{\varphi}W^I_{\mu\nu},
    &
    O_{tG}=g_sy_t(\bar{Q}\sigma^{\mu\nu}T^At)\tilde{\varphi}G^A_{\mu\nu},
    \\
    O_{t\varphi}=y_t^3(\varphi^\dagger\varphi)(\bar{Q}t)\tilde\varphi,
    &
    O_{\varphi G}=y_t^2\left(\varphi^\dagger\varphi\right)\left(
    G_{\mu\nu}^AG^{A\mu\nu} \right),
    \nonumber
  \end{array}
  \label{eq:allfc}
\end{equation}
where $g_{s,W,Y}$ are the SM gauge couplings, and $y_t$ is the top Yukawa
coupling. $Q$ is the third generation left-handed quark doublet, while
$t$ is the right-handed top quark.  The last operator $O_{\varphi G}$ does not
involve a top quark, but
we include it here because it contributes to $t\bar tH$ production as
well as several top-loop induced processes.  Furthermore, it has
$\mathcal{O}(\alpha_s)$ mixing terms to $O_{t\varphi}$ and from $O_{tG}$, and
so should not be ignored, in particular if the latter is present.  Four-fermion
operators, on the other hand, should also be incorporated, but we do not list
them because of their huge number.  For a discussion on four-fermion operators
see \cite{AguilarSaavedra:2010zi}.  Finally, the operators $O_{\varphi\varphi}$
and $O_{bW}$ are not included as they do not contribution at order
$\Lambda^{-2}$ in the $m_b\to0$ limit.

Our convention for the normalization of the operators are chosen such that
they do not modify the counting of the QCD and electroweak orders comparing
with the corresponding SM vertex.  It is also consistent with the naive
dimension analysis
\cite{Manohar:1983md}, except that $C_{\varphi G}$ is suppressed by a loop
factor, $\alpha_s/4\pi$.  Under this convention, in Table~1 we mark the current
status of all $\mathcal{O}(\Lambda^{-2})$ contributions that exist at the
leading QCD order, either at leading order (LO), or at NLO in QCD, or via a
top-quark loop, depending on whether they are analytically available, or
available with {\sc MG5\_aMC} for event generation.  Note, that we did not
define the coupling order of the four-fermion operator coefficients.  For
these operators, in Table~1 we only show contributions that are present at the
LO, and correspond to either $\mathcal{O}\left(\frac{C_{4f}}{\Lambda^2
g_s^2}\right)$ or $\mathcal{O}\left(\frac{C_{4f}}{\Lambda^2 g_{W,Y}^2}\right)$
corrections to the SM, if these operators are not defined with any prefactors.

\begin{table}
  \newcommand{\amark}{a}
  \newcommand{\camark}{A}
  \newcommand{\pmark}{P}
  \begin{center}
    \begin{tabular}{lcccccccccc}
      \hline\hline
      Process & $O_{tG}$ & $O_{tB}$ & $O_{tW}$ & $O_{\varphi Q}^{(3)}$ &
      $O_{\varphi Q}^{(1)}$ & $O_{\varphi t}$ & $O_{t\varphi}$ &
      $O_{\mbox{4f}}$ & $O_{\varphi G}$
      \\ \hline
      $t\to bW\to bl^+\nu$
      &\amark& &\amark & \amark &&&& \amark & \\
      $pp\to t\bar q$ &\camark & &\camark & \camark & & & &\camark & 
      \\
      $pp\to tW$
      &\camark & &\camark & \camark &&&&&
      \\
      $pp\to t\bar t$ 
      &\camark&&&&&&&\pmark&
      \\
      $pp\to t\bar t\gamma$
      &\camark &\camark &\camark & & & &&\pmark&
      \\
      $pp\to t \gamma j$
      &\camark &\camark &\camark &\camark & & &&\pmark&
      \\
      $pp\to t\bar tZ$
      &\camark &\camark &\camark &\camark &\camark &\camark &&\pmark&
      \\
      $pp\to t Zj$
      &\camark &\camark &\camark &\camark &\camark &\camark &&\pmark&
      \\
      $pp\to t\bar tW$
      &\camark & & &&& &&\pmark&
      \\
      $e^+e^-\to t\bar t$
      &\camark &\camark &\camark &\camark&\camark&\camark &&\camark&
      \\
      $pp\to t\bar tH$
      &\pmark & & &&& &\camark&\pmark&\pmark
      \\
      $pp\to tHj$
      &\pmark & &\pmark &\pmark&& &\camark&\pmark&\pmark
      \\
      {$gg\to H, Hj$}
      &\camark & & &&& &\camark&&\camark
      \\
      {$gg\to HZ$}
      &\camark & & &\camark&\camark&\camark &\camark&&\camark
      \\\hline\hline
    \end{tabular}
  \caption{Top-quark operators and key processes at the LHC, for the flavor-diagonal
	  sector.  Contributions which occur at the leading QCD order either at
	  LO, or NLO, or through a top-quark loop, is marked by ``a'', ``A'' or
	  ``P''.  ``a'': result is known analytically; ``A'': NLO simulation
	  within {\sc MG5\_aMC} is available;  ``P'': NLO simulation is not
	  available currently, but is planned.  $O_{4f}$ denotes any
	  four-fermion operator.  \label{tab:fd} }
  \end{center}
\end{table}

\begin{figure}[htb]
	\begin{center}
		\includegraphics[width=.48\linewidth]{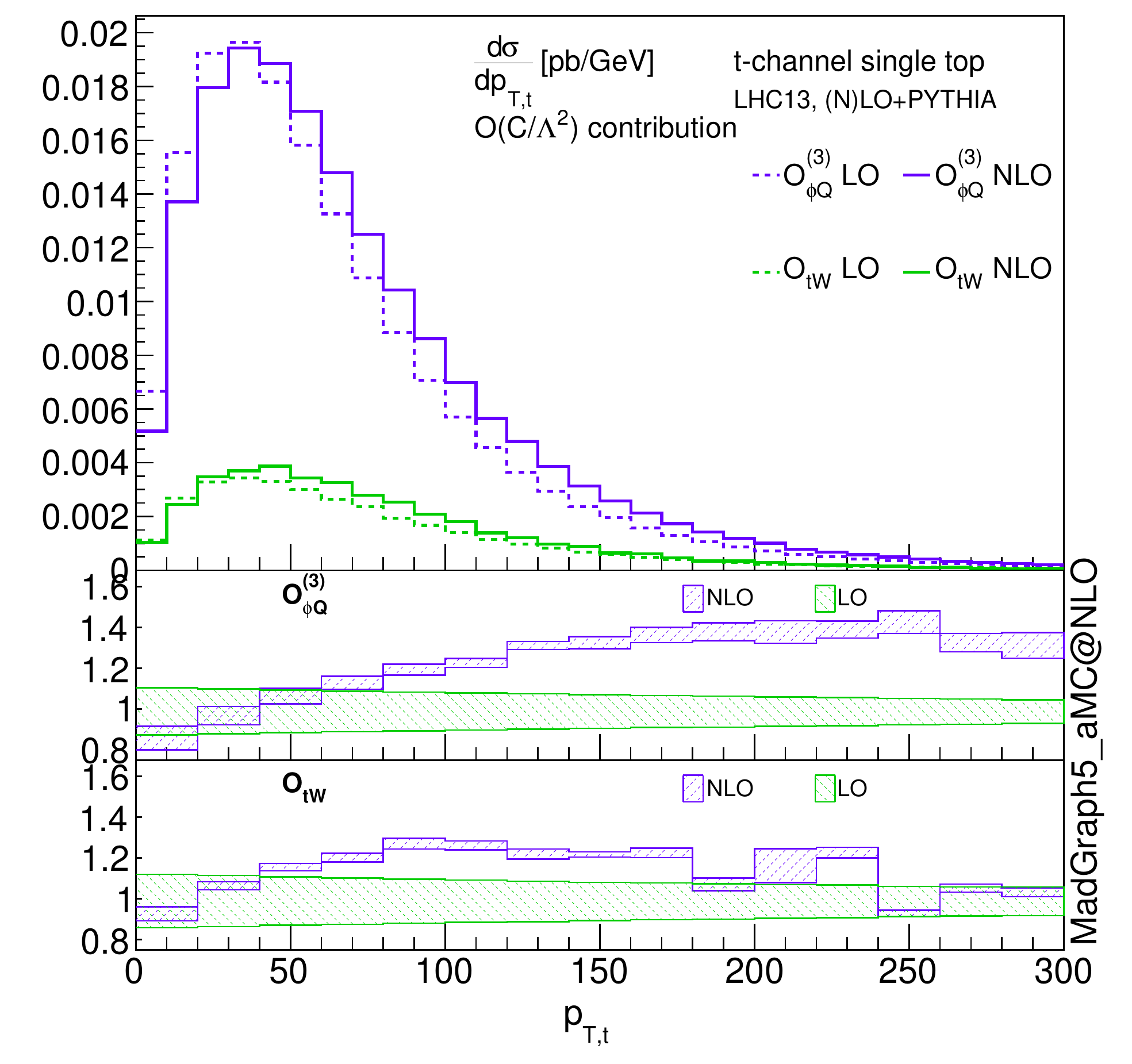}
		\includegraphics[width=.48\linewidth]{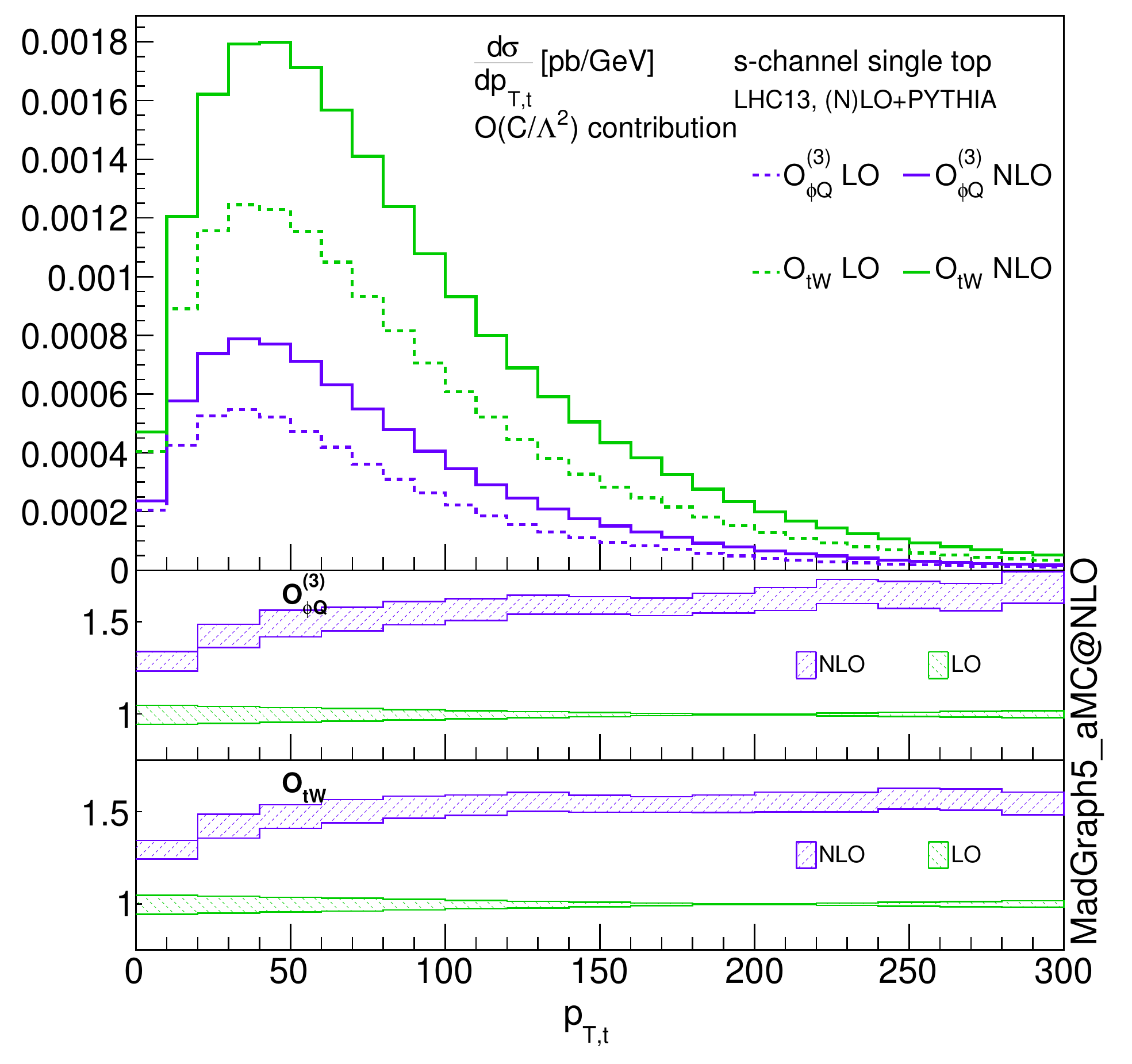}
	\end{center}
	\caption{$p_T$ distributions of the top quark, in single top production
		at the LHC 13 TeV, from the operators $O_{\varphi Q}^{(3)}$
		and $O_{tW}$.  Left: $t$-channel. Right: $s$-channel.  Only
	the interferences between the SM and the operators are displayed.
	Anti-top quark is not included.
        $K$-factors and scale uncertainties are given in the lower panel.}
	\label{fig:ptt}
\end{figure}

As we can see from Table 1, apart from the decay process which is analytically known
\cite{Zhang:2014rja},
the full set of operators that represent deviations in the top-quark couplings
to SM gauge bosons can be already simulated at NLO in QCD, for processes that are
relevant for their measurements.  The chromo-dipole operator has been implemented
with $t\bar t$ production in Ref.~\cite{Franzosi:2015osa}.  All
the other operators involving the top-quark and the gauge-boson fields have
been implemented in recent works,
focusing on single top processes \cite{t} and on $t\bar tV$ processes and
$e^+e^-\to t\bar t$ \cite{ttV}.  In these works the corresponding counterterms
are computed with the help of the {\sc NLOCT} package and by using the anomalous
dimension matrix for the dipole operators, which is available from
Ref.~\cite{Zhang:2014rja}.  Some results for single top production (in
the leptonic channel) are shown in Figure~\ref{fig:ptt} for illustration,
where the interference contributions from $O_{\varphi Q}^{(3)}$ and $O_{tW}$ are
compared at LO and NLO.  The same approach also covers the loop-induced process
initiated by two gluons as well as $t\bar t$ production at lepton colliders.
Also available is the top-Yukawa operator $O_{t\varphi}$ in $t\bar tH$ and $tH$
production, via the Higgs Characterisation (HC) model
\cite{Artoisenet:2013puc,Demartin:2014fia,Demartin:2015uha}.  The other
operators relevant for these two processes are now being studied. Finally,
implementation of four-fermion operators is planned.  Note that the
four-fermion operator that enters the single top production process can be
simulated by introducing a heavy $W'$ propagator that mimics the contact interaction,
because the QCD corrections to the two fermion currents factorize.
Corresponding counterterms can be computed with {\sc NLOCT} which is publicly
available. The similar trick can be used also in $e^+e^-\to t\bar t$.

We now turn to the flavor-changing sector. Deviations in $tqB$ vertex, where $B$ is
either a gauge boson or a Higgs boson, are captured by the following operators
\begin{equation}
  \begin{array}{lll}
    O_{\varphi q}^{(3,i+3)}=i\FDFI(\bar{q}_i\gamma^\mu\tau^IQ)\,,
    &
    O_{\varphi q}^{(1,i+3)}=i\FDF(\bar{q}_i\gamma^\mu Q) \,,
    \\
    O_{\varphi u}^{(i+3)}=i\FDF(\bar{u}_i\gamma^\mu t)\,,
    &
    O_{uB}^{(i3)}=g_Yy_t(\bar{q}_i\sigma^{\mu\nu}t)\tilde{\varphi}B_{\mu\nu}\,,
    \\
    O_{uW}^{(i3)}=g_Wy_t(\bar{q}_i\sigma^{\mu\nu}\tau^It)\tilde{\varphi}W^I_{\mu\nu}\,,
    &
    O_{uG}^{(i3)}=g_sy_t(\bar{q}_i\sigma^{\mu\nu}T^At)\tilde{\varphi}G^A_{\mu\nu}\,,
    \\
    O_{u\varphi}^{(i3)}=y_t^3(\varphi^\dagger\varphi)(\bar{q}_it)\tilde\varphi\,,
    \nonumber
  \end{array}
  \label{eq:allfcnc}
\end{equation}
where $i=1,2$ is the flavor index.  For operators with $(i3)$ superscript,
a similar set of operators with $(3i)$ flavor structure can be obtained
by interchanging $(i3)\leftrightarrow(3i)$, $t\leftrightarrow u_i$,
and $Q\leftrightarrow q_i$.  In addition, similar to the flavor-diagonal
sector, four-fermion operators could also be relevant for the key processes.  

The status of these operators in all the relevant processes are outlined in
Table~2, similar to Table~1.  The three flavor-changing decay channels are
known in Ref.~\cite{Zhang:2013xya,Zhang:2014rja}. 
The single top production channels are automated with {\sc MG5\_aMC}
\cite{Degrande:2014tta}, and the corresponding NLO UFO model is available at
the {\sc FeynRules} repository \cite{fr}.  The relevant four-fermion operators
involve a top quark, a light quark, and two leptons.  They are not currently
implemented, but are planned to be, and in fact most of them can be easily
done by introducing a heavy mediator, with the help of NLOCT, except for
$O_{lequ}^{(3)}$ which has a tensor
structure.  In Ref.~\cite{Durieux:2014xla} we have made use of these results
and performed a global fit, for all flavor-changing operators.  Even though the
fit is based on several simplifications, it provides a proof of principle that
a global fitting program can be carried out entirely with NLO accuracy.

\begin{table}
  \newcommand{\amark}{a}
  \newcommand{\camark}{A}
  \newcommand{\pmark}{P}
  \begin{center}
    \begin{tabular}{lcccccccc}
      \hline\hline
      Process & $O_{\phi q}^{(3)}$ & $O_{\phi q}^{(1)}$ & $O_{\phi u}^{(1)}$ 
      & $O_{uW}$ & $O_{uB}$ & $O_{uG}$ &$O_{u\phi}$ & $O_{\mbox{4f}}$
      \\\hline
      $t\to qZ^*,q\gamma^*\to ql^+l^-$
      &\amark& \amark &\amark & \amark &\amark&\amark&&\amark \\
      $t\to q\gamma$
      &&  & & \amark &\amark&\amark&& \\
      $t\to qH$
      &&  & &  &&\amark&\amark& \\
      $pp\to t$
      &&  & &  &&\camark&& \\
      $pp\to tZ^*,t\gamma^*\to tl^+l^-$
      &\camark& \camark &\camark & \camark &\camark&\camark&&(\camark) \\
      $pp\to t\gamma$
      &&  & & \camark &\camark&\camark&& \\
      $pp\to tH$
      &&  & &  &&\camark&\camark& 
      \\\hline\hline
    \end{tabular}
    \caption{Similar to Table~1 but for the flavor-changing sector, and
	    contributions are at $\mathcal{O}(\Lambda^{-4})$.
 The flavor indices of operators are omitted.}
  \end{center}
\end{table}

\section{The Higgs and electroweak sector}

\begin{figure}[htb]
	\begin{center}
		\includegraphics[width=.6\linewidth]{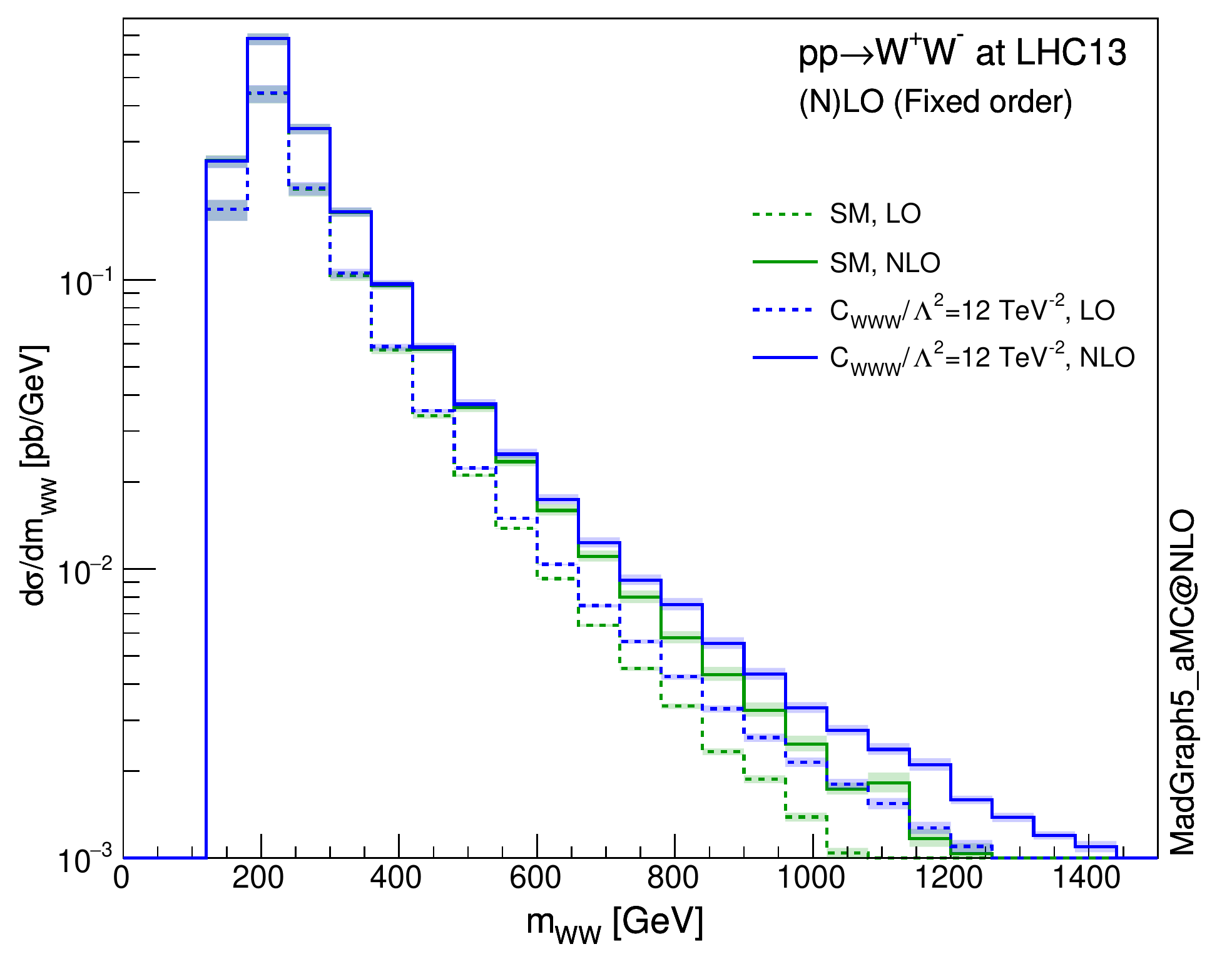}
	\end{center}
	\caption{Invariant mass distribution of $W$-boson pair, at the LHC
		13 TeV, in the SM and with the operator $O_{WWW}$.}
	\label{fig:mww}
\end{figure}

In the electroweak sector, operators that do not include any
colored field are trivial to add at NLO in QCD, as they do not generate any new
dimension-six counterterms.  For example, the dimension-six triple gauge-boson
operators, the dimension-eight quartic gauge-boson operators, and the operators
involving the Higgs boson and the SM weak gauge bosons, are straightforward
with the help
of {\sc NLOCT}.  As a result, processes such as $W$-pair production and vector-boson
scattering can be simulated at NLO in QCD, with deviations in the triple or the
quartic gauge-boson couplings.
For illustration, in Figure~\ref{fig:mww} we show the
invariant mass distribution of the $W$-pair production with the operator
$O_{WWW}$ (see, for example, Ref.~\cite{Degrande:2012wf} for its definition),
both at LO and NLO.
Higgs production processes in vector-boson fusion and in $VH$ associated
channel are covered in a similar way, with higher-dimensional interactions
between the Higgs and electroweak gauge bosons (see also
\cite{Maltoni:2013sma}).  

Some Higgs operators involve colored fields.  For Higgs measurements the most
important ones are $O_{\varphi G}$ and $O_{t\varphi}$.  With a slightly
different convention, these operators are available in the HC framework,
through which all the main Higgs production channels
(gluon fusion, weak vector-boson fusion and associated production, and $t\bar tH$)
\cite{Maltoni:2013sma,Demartin:2014fia}, as well as the subdominant process,
associated production with a single top quark \cite{Demartin:2015uha},
have been studied.  The UFO model is available at the {\sc FeynRules} repository
\cite{hc}.  In addition, the Higgs Effective Lagrangian (HEL), described in
Ref.~\cite{Alloul:2013naa}, is now being extended to NLO in QCD \cite{hel}.

\section{Summary}

NLO simulation for the SM EFT has started to become available with the {\sc
MG5\_aMC} framework.  Even though the NLO implementation of operators is not
fully automated at the moment, in several case by case studies, some UFO models
have been made available.  With these models, cross sections and differential
distributions matched to parton shower can be computed up to NLO in QCD, in a
flexible and fully automatic way, just like in the SM.  We have briefly
discussed several works where such implementations for certain processes and
operators are provided.  Remarkably, the most important operators in the top
quark, the electroweak, and the Higgs sectors are either already available or
under investigation.  These studies improve the SM predictions at
dimension-six, both in accuracy and in precision, and pave the way towards an
accurate global fit for SM deviations.

\acknowledgments
I would like to thank F.~Maltoni for constantly supporting projects on EFT at
NLO in QCD.  I am grateful for the collaborations with C.~Degrande,
O.~B.~Bylund, D.~B.~Franzosi, I.~Tsinikos, E.~Vryonidou, and J.~Wang on various
top-EFT projects.  I would also like to thank F.~Demartin,
K.~Mawatari, and K.~Mimasu for informing the status of HC and HEL.  The work of
C.Z.~is supported by U.S.~Department of Energy under Grant DE-SC0012704.


\begin{thebibliography}{99}

\bibitem{Buchmuller:1985jz} 
  W.~Buchmuller and D.~Wyler,
  \emph{Effective Lagrangian Analysis of New Interactions and Flavor Conservation,}
  \emph{Nucl.\ Phys.\ B }{\bf 268}, 621 (1986).
  %doi:10.1016/0550-3213(86)90262-2
  %%CITATION = doi:10.1016/0550-3213(86)90262-2;%%
  %1080 citations counted in INSPIRE as of 15 Jan 2016

\bibitem{Agashe:2014kda}
  K.~A.~Olive {\it et al.} [Particle Data Group Collaboration],
  \emph{Review of Particle Physics,}
  \emph{Chin.\ Phys.\ C }{\bf 38}, 090001 (2014).
  %doi:10.1088/1674-1137/38/9/090001
  %%CITATION = doi:10.1088/1674-1137/38/9/090001;%%

\bibitem{Alwall:2014hca} 
  J.~Alwall {\it et al.},
  \emph{The automated computation of tree-level and next-to-leading order differential cross sections, and their matching to parton shower simulations,}
  \emph{JHEP }{\bf 1407}, 079 (2014)
  %doi:10.1007/JHEP07(2014)079
  [{\tt arXiv:1405.0301 [hep-ph]}].
  %%CITATION = doi:10.1007/JHEP07(2014)079;%%
  %693 citations counted in INSPIRE as of 14 Jan 2016

\bibitem{Frixione:2002ik} 
  S.~Frixione and B.~R.~Webber,
  \emph{Matching NLO QCD computations and parton shower simulations,}
  \emph{JHEP }{\bf 0206}, 029 (2002)
  %doi:10.1088/1126-6708/2002/06/029
  [{\tt hep-ph/0204244}].
  %%CITATION = doi:10.1088/1126-6708/2002/06/029;%%
  %1990 citations counted in INSPIRE as of 14 Jan 2016

\bibitem{Alloul:2013bka} 
  A.~Alloul, N.~D.~Christensen, C.~Degrande, C.~Duhr and B.~Fuks,
  \emph{FeynRules  2.0 - A complete toolbox for tree-level phenomenology,}
  \emph{Comput.\ Phys.\ Commun.\  }{\bf 185}, 2250 (2014)
  %doi:10.1016/j.cpc.2014.04.012
  [{\tt arXiv:1310.1921 [hep-ph]}].
  %%CITATION = doi:10.1016/j.cpc.2014.04.012;%%
  %324 citations counted in INSPIRE as of 14 Jan 2016

\bibitem{Degrande:2011ua} 
  C.~Degrande, C.~Duhr, B.~Fuks, D.~Grellscheid, O.~Mattelaer and T.~Reiter,
  \emph{UFO - The Universal FeynRules Output,}
  \emph{Comput.\ Phys.\ Commun.\  }{\bf 183}, 1201 (2012)
  %doi:10.1016/j.cpc.2012.01.022
  [{\tt arXiv:1108.2040 [hep-ph]}].
  %%CITATION = doi:10.1016/j.cpc.2012.01.022;%%
  %266 citations counted in INSPIRE as of 14 Jan 2016

\bibitem{Hirschi:2011pa} 
  V.~Hirschi, R.~Frederix, S.~Frixione, M.~V.~Garzelli, F.~Maltoni and R.~Pittau,
  \emph{Automation of one-loop QCD corrections,}
  \emph{JHEP }{\bf 1105}, 044 (2011)
  %doi:10.1007/JHEP05(2011)044
  [{\tt arXiv:1103.0621 [hep-ph]}].
  %%CITATION = doi:10.1007/JHEP05(2011)044;%%
  %246 citations counted in INSPIRE as of 14 Jan 2016

\bibitem{Degrande:2014vpa} 
  C.~Degrande,
  \emph{Automatic evaluation of UV and R2 terms for beyond the Standard Model Lagrangians: a proof-of-principle,}
  \emph{Comput.\ Phys.\ Commun.\  }{\bf 197}, 239 (2015)
  %doi:10.1016/j.cpc.2015.08.015
  [{\tt arXiv:1406.3030 [hep-ph]}].
  %%CITATION = doi:10.1016/j.cpc.2015.08.015;%%
  %24 citations counted in INSPIRE as of 14 Jan 2016

\bibitem{AguilarSaavedra:2008zc} 
  J.~A.~Aguilar-Saavedra,
  \emph{A Minimal set of top anomalous couplings,}
  \emph{Nucl.\ Phys.\ B }{\bf 812}, 181 (2009)
  %doi:10.1016/j.nuclphysb.2008.12.012
  [{\tt arXiv:0811.3842 [hep-ph]}].
  %%CITATION = doi:10.1016/j.nuclphysb.2008.12.012;%%
  %201 citations counted in INSPIRE as of 14 Jan 2016

\bibitem{AguilarSaavedra:2009mx} 
  J.~A.~Aguilar-Saavedra,
  \emph{A Minimal set of top-Higgs anomalous couplings,}
  \emph{Nucl.\ Phys.\ B }{\bf 821}, 215 (2009)
  %doi:10.1016/j.nuclphysb.2009.06.022
  [{\tt arXiv:0904.2387 [hep-ph]}].
  %%CITATION = doi:10.1016/j.nuclphysb.2009.06.022;%%
  %95 citations counted in INSPIRE as of 14 Jan 2016

\bibitem{AguilarSaavedra:2010zi} 
  J.~A.~Aguilar-Saavedra,
  \emph{Effective four-fermion operators in top physics: A Roadmap,}
  \emph{Nucl.\ Phys.\ B }{\bf 843}, 638 (2011),
  \emph{Nucl.\ Phys.\ B }{\bf 851}, 443 (2011)
  %doi:10.1016/j.nuclphysb.2011.06.003, 10.1016/j.nuclphysb.2010.10.015
  [{\tt arXiv:1008.3562 [hep-ph]}].
  %%CITATION = doi:10.1016/j.nuclphysb.2011.06.003, 10.1016/j.nuclphysb.2010.10.015;%%
  %84 citations counted in INSPIRE as of 14 Jan 2016

\bibitem{Manohar:1983md} 
  A.~Manohar and H.~Georgi,
  \emph{Chiral Quarks and the Nonrelativistic Quark Model,}
  \emph{Nucl.\ Phys.\ B }{\bf 234}, 189 (1984).
  %doi:10.1016/0550-3213(84)90231-1
  %%CITATION = doi:10.1016/0550-3213(84)90231-1;%%
  %1724 citations counted in INSPIRE as of 14 Jan 2016

\bibitem{Zhang:2014rja} 
  C.~Zhang,
  \emph{Effective field theory approach to top-quark decay at next-to-leading order in QCD,}
  \emph{Phys.\ Rev.\ D }{\bf 90}, no. 1, 014008 (2014)
  %doi:10.1103/PhysRevD.90.014008
  [{\tt arXiv:1404.1264 [hep-ph]}].
  %%CITATION = doi:10.1103/PhysRevD.90.014008;%%
  %12 citations counted in INSPIRE as of 14 Jan 2016

\bibitem{Franzosi:2015osa} 
  D.~Buarque Franzosi and C.~Zhang,
  \emph{Probing the top-quark chromomagnetic dipole moment at next-to-leading order in QCD,}
  \emph{Phys.\ Rev.\ D }{\bf 91}, no. 11, 114010 (2015)
  %doi:10.1103/PhysRevD.91.114010
  [{\tt arXiv:1503.08841 [hep-ph]}].
  %%CITATION = doi:10.1103/PhysRevD.91.114010;%%
  %5 citations counted in INSPIRE as of 14 Jan 2016

\bibitem{t} 
  C.~Zhang,
  \emph{in progress.}
\bibitem{ttV} 
  O.~B.~Bylund F.~Maltoni, I.~Tsinikos, E.~Vryonidou and C.~Zhang,
  \emph{in progress.}

\bibitem{Artoisenet:2013puc} 
  P.~Artoisenet {\it et al.},
  \emph{A framework for Higgs characterisation,}
  \emph{JHEP }{\bf 1311}, 043 (2013)
  %doi:10.1007/JHEP11(2013)043
  [{\tt arXiv:1306.6464 [hep-ph]}].
  %%CITATION = doi:10.1007/JHEP11(2013)043;%%
  %62 citations counted in INSPIRE as of 14 Jan 2016

\bibitem{Demartin:2014fia} 
  F.~Demartin, F.~Maltoni, K.~Mawatari, B.~Page and M.~Zaro,
  \emph{Higgs characterisation at NLO in QCD: CP properties of the top-quark Yukawa interaction,}
  \emph{Eur.\ Phys.\ J.\ C }{\bf 74}, no. 9, 3065 (2014)
  %doi:10.1140/epjc/s10052-014-3065-2
  [{\tt arXiv:1407.5089 [hep-ph]}].
  %%CITATION = doi:10.1140/epjc/s10052-014-3065-2;%%
  %30 citations counted in INSPIRE as of 14 Jan 2016

\bibitem{Demartin:2015uha} 
  F.~Demartin, F.~Maltoni, K.~Mawatari and M.~Zaro,
  \emph{Higgs production in association with a single top quark at the LHC,}
  \emph{Eur.\ Phys.\ J.\ C }{\bf 75}, no. 6, 267 (2015)
  %doi:10.1140/epjc/s10052-015-3475-9
  [{\tt arXiv:1504.00611 [hep-ph]}].
  %%CITATION = doi:10.1140/epjc/s10052-015-3475-9;%%
  %11 citations counted in INSPIRE as of 14 Jan 2016

\bibitem{Zhang:2013xya} 
  C.~Zhang and F.~Maltoni,
  \emph{Top-quark decay into Higgs boson and a light quark at next-to-leading order in QCD,}
  \emph{Phys.\ Rev.\ D }{\bf 88}, 054005 (2013)
  %doi:10.1103/PhysRevD.88.054005
  [{\tt arXiv:1305.7386 [hep-ph]}].
  %%CITATION = doi:10.1103/PhysRevD.88.054005;%%
  %23 citations counted in INSPIRE as of 14 Jan 2016

\bibitem{Degrande:2014tta} 
  C.~Degrande, F.~Maltoni, J.~Wang and C.~Zhang,
  \emph{Automatic computations at next-to-leading order in QCD for top-quark flavor-changing neutral processes,}
  \emph{Phys.\ Rev.\ D }{\bf 91}, 034024 (2015)
  %doi:10.1103/PhysRevD.91.034024
  [{\tt arXiv:1412.5594 [hep-ph]}].
  %%CITATION = doi:10.1103/PhysRevD.91.034024;%%
  %4 citations counted in INSPIRE as of 14 Jan 2016

\bibitem{fr}
  \url{http://feynrules.irmp.ucl.ac.be/wiki/TopFCNC}

\bibitem{Durieux:2014xla} 
  G.~Durieux, F.~Maltoni and C.~Zhang,
  \emph{Global approach to top-quark flavor-changing interactions,}
  \emph{Phys.\ Rev.\ D }{\bf 91}, no. 7, 074017 (2015)
  %doi:10.1103/PhysRevD.91.074017
  [{\tt arXiv:1412.7166 [hep-ph]}].
  %%CITATION = doi:10.1103/PhysRevD.91.074017;%%
  %12 citations counted in INSPIRE as of 14 Jan 2016

\bibitem{Degrande:2012wf} 
  C.~Degrande, N.~Greiner, W.~Kilian, O.~Mattelaer, H.~Mebane, T.~Stelzer, S.~Willenbrock and C.~Zhang,
  \emph{Effective Field Theory: A Modern Approach to Anomalous Couplings,}
  \emph{Annals Phys.\  }{\bf 335}, 21 (2013)
  %doi:10.1016/j.aop.2013.04.016
  [{\tt arXiv:1205.4231 [hep-ph]}].
  %%CITATION = doi:10.1016/j.aop.2013.04.016;%%
  %48 citations counted in INSPIRE as of 14 Jan 2016

\bibitem{Maltoni:2013sma} 
  F.~Maltoni, K.~Mawatari and M.~Zaro,
  \emph{Higgs characterisation via vector-boson fusion and associated production: NLO and parton-shower effects,}
  \emph{Eur.\ Phys.\ J.\ C }{\bf 74}, no. 1, 2710 (2014)
  %doi:10.1140/epjc/s10052-013-2710-5
  [{\tt arXiv:1311.1829 [hep-ph]}].
  %%CITATION = doi:10.1140/epjc/s10052-013-2710-5;%%
  %19 citations counted in INSPIRE as of 14 Jan 2016

\bibitem{hc}
	\url{http://feynrules.irmp.ucl.ac.be/wiki/HiggsCharacterisation}

\bibitem{Alloul:2013naa} 
  A.~Alloul, B.~Fuks and V.~Sanz,
  \emph{Phenomenology of the Higgs Effective Lagrangian via FEYNRULES,}
  \emph{JHEP }{\bf 1404}, 110 (2014)
  %doi:10.1007/JHEP04(2014)110
  [{\tt arXiv:1310.5150 [hep-ph]}].
  %%CITATION = doi:10.1007/JHEP04(2014)110;%%
  %45 citations counted in INSPIRE as of 14 Jan 2016

\bibitem{hel} 
  C.~Degrande, B.~Fuks, K.~Mawatari, K.~Mimasu and V.~Sanz,
  \emph{in progress.}
\end{thebibliography}
\end{document}